\input harvmac
\input epsf
\noblackbox
\Title{\vbox{\baselineskip12pt\hbox{UCLA/96/TEP/30}\hbox{hep-th/9609132}}}
{\vbox{\centerline{Note on the Off-Shell Equivalence between}
\vskip2pt\centerline{the Linear and Non-Linear Sigma Models$^\star$}}}
\footnote{}{$^\star$ This work was supported in part by the U.S.
Department of Energy, under Contract DE-AT03-88ER 40384 Mod A006 Task C.}

\centerline{Hidenori SONODA$^\dagger$\footnote{}{$^\dagger$
E-mail: sonoda@physics.ucla.edu; Address after 1 January 1996:
Department of Physics, Kobe University, Kobe 657, Japan.}}
\bigskip\centerline{\it Department of Physics and Astronomy, UCLA,
Los Angeles, CA 90095-1547, USA}

\vskip 1in
As is well known, one can arrange the parameters of the O(N)
non-linear sigma model to reproduce the low energy S-matrix
elements of the renormalizable O(N) linear sigma model.  In this
note we provide details which are necessary in order to obtain
the off-shell equivalence between the two theories.

\Date{September 1996}

\def\no{\noindent}
\def\vev#1{\left\langle #1 \right\rangle}
\def\ep{\epsilon}
\def\L{{\cal L}}
\def\dt{{d \over dt}}
\def\g{g_{-2}}
\def\dots{{~...~}}
\def\e{\hbox{e}}

\newsec{Introduction and summary}

The idea of low energy effective theories
is simple: phenomena at energy less than a scale
$\Lambda$ can be described by a field theory which only contains
particles of mass less than $\Lambda$.  The most notable example is
the chiral lagrangian for the low energy pion physics
\ref\rweinberg{S.~Weinberg, Phys.~Rev.~{\bf 166}(1968)1568;
Physica~{\bf 96A}(1979)327}.  The chiral lagrangian
is given in terms of the pion fields such that it is
invariant under the flavor SU(2)$\times$SU(2).  At tree level (i.e.,
in the soft pion limit)
the lagrangian has only one parameter $f_\pi$, the
pion decay constant, and we obtain all the consequences of current algebra
simply by expanding the lagrangian and reading out the matrix elements.
Away from the soft pion limit, higher
order corrections have been calculated in order to achieve better fits
with experiments \ref\rgl{J.~Gasser and H.~Leutwyler,
Annl.~Phys.~{\bf 158}(1984)142}.
As far as S-matrix elements are concerned, the procedure
of loop expansions is well understood.

The purpose of this note is to give a clear statement
of the off-shell equivalence between the O(N) linear
and non-linear sigma models.  The elementary fields of
the linear sigma model are unconstrained scalar fields
$\phi^I~(I=1,\dots,N)$ which transform as a vector of O(N).  On
the other hand the elementary fields of the non-linear sigma
model are the vector $\Phi^I$ which are constrained by the
non-linear relation
\eqn\enonlinear{\sum_{I=1}^N \Phi^I \Phi^I = 1~.}
The off-shell equivalence means that at energies lower than
the symmetry breaking scale,
the correlation functions of $\phi$'s in the linear sigma
model can be reproduced by certain correlation functions in the
non-linear sigma model.

Let us briefly summarize the main points of this note.  First we
give two points on the renormalization properties of
the non-linear sigma model:

\no
(A) The field $\Phi^I$ which transforms linearly under O(N) mixes with
higher dimensional fields which also transform as vectors of O(N).

\no
(B) In considering the Fourier transforms of the correlation functions,
the field $\Phi^I$ is not renormalized linearly: we must introduce
counterterms to remove the ultraviolet (UV) divergences due to
two or more $\Phi$'s at the same point in space.

The first point is nothing new.  Since the non-linear sigma model is
non-renormalizable, the field $\Phi^I$ cannot be renormalized
multiplicatively, but it mixes with an infinite number of
higher dimensional fields which transform
as vectors under O(N).  The second point, which is familiar from the
renormalization of composite fields in renormalizable field theories
(see, for example, ref.~\ref\rcollins{J.~Collins,
Renormalization (Cambridge University Press, 1984)}),
has not been emphasized before in the literature.  This point is not
only important for the removal of UV divergences but also for obtaining
the off-shell equivalence.  See (B$^\prime$) below.

Similarly, we give two points on the off-shell equivalence:

\no
(A$^\prime$) The elementary field $\phi^I$ of
the linear sigma model corresponds
to a linear combination of the fields in the non-linear sigma model all of
which transform as vectors of O(N).  More specifically, $\phi^I$ corresponds
to a linear combination of the non-linear field $\Phi^I$ and fields with
two or more derivatives.

\no
(B$^\prime$) For the off-shell equivalence, we must also introduce finite
counterterms corresponding to two or more fields at the same point
in space.

The loop expansion of the O(N) non-linear sigma model was
discussed long ago by Appelquist and Bernard
\ref\rab{T.~Appelquist and C.~Bernard, Phys.~Rev.~{\bf
D23}(1981)425} and by Akhoury and Yao \ref\ray{R.~Akhoury and Y.-P.~Yao,
Phys.~Rev.~{\bf D25}(1982)3361}.  The results of this note are refinements
of some of the results of the above mentioned works.

The present paper is organized as follows.  In sect.~2, we summarize the
one-loop corrections to the two- and four-point correlation functions
of $\Phi$'s
in the O(N) non-linear sigma model.  We only give enough details to clarify our
points (A,B) above.  In sect.~3, we consider the one-loop corrections
to the two- and four-point correlation functions of $\phi$'s in the O(N)
linear sigma model, and take their low energy approximations.
In sect.~4 we compare the results of sects.~2 and 3, and determine
the parameters of the low energy effective theory to obtain the off-shell
equivalence.  In sect.~5 we give concluding remarks.

Throughout this note we will use the dimension regularization with
\eqn\eD{D \equiv 4 - \ep~.}
For simplicity, we will use the euclidean metric.

\newsec{The loop expansion of the O(N) non-linear sigma model}

The O(N) non-linear sigma model is defined by the lagrangian
\eqn\enls{\eqalign{&\L_{eff} =
{1 \over 2 \g} \partial_\mu \Phi^I \partial_\mu \Phi^I \cr
& + {1 \over \g^2} \Big( ~G_{-4}^{(1)}
{}~(\partial_\mu \Phi^I \partial_\mu \Phi^I)^2
+ G_{-4}^{(2)}
{}~\partial_\mu \Phi^I \partial_\nu \Phi^I
\partial_\mu \Phi^J \partial_\nu \Phi^J \cr
&\qquad\qquad + G_{-4}^{(3)}
{}~\partial^2 \Phi^I \partial^2 \Phi^I ~\Big)
 + \dots,\cr}}
where the field $\Phi^I~(I=1,\dots,N)$ takes values on the unit
sphere $S^N$, and the summation over the repeated vector indices
$I,J$ is implied.  The dots represent terms with six or more
derivatives.  There are only three terms possible with four
derivatives.  The three parameters are given as
\eqn\eG{G_{-4}^{(i)} = {z^{(i)} \over \ep}~\g^2 + g_{-4}^{(i)}~,}
where $z^{(i)}$, $g_{-4}^{(i)}$ are finite constants.  The
constants $z^{(i)}$ are chosen to remove the one-loop UV
divergences.  The finite parameter $g_{-4}^{(i)}$ satisfies the
renormalization group (RG) equation:
\eqn\egfourRG{\dt g_{-4}^{(i)} = (-4+\ep) g_{-4}^{(i)} - z^{(i)} \g^2~.}

Before discussing the
the renormalization of the field $\Phi^I$,
we must make an important remark.  The fields $\Phi^I~(I=1,\dots,N)$ satisfy
the non-linear constraint \enonlinear.  Given the convention
\eqn\evevNL{\vev{\Phi^i} = 0\quad (i=1,\dots,N-1)~,}
a standard parametrization is
\eqn\epi{\Phi^i = \sqrt{\g}~\pi^i~(i=1,\dots,N-1)~,\quad
\Phi^N = \sqrt{1 - \g~\vec{\pi} \cdot \vec{\pi}}~,}
where $\pi^i$ are ${\rm N-1}$ independent fields.  But what parametrization
we use is irrelevant as long as we use the dimension regularization, since
the jacobian of the infinitesimal field transformation
\eqn\etrans{\eqalign{&\pi^i \to (1 + \ep_0 (\pi^2)) \pi^i +
\ep_1^{ijk} (\pi^2) \partial_\mu \pi^j \partial_\mu \pi^k
+ \ep_2^{ij} (\pi^2) \partial^2 \pi^j \cr
&\qquad\qquad + (\hbox{higher~order~derivatives})\cr}}
is unity.  (See ref.~\ref\rarzt{C.~Arzt, Phys.~Lett.~{\bf B342}(1995)189}
for a recent discussion.)  Therefore, we need not discuss the fields
$\pi^i$ which give a non-linear realization of O(N), but we only need to
discuss linear representations of O(N) such as $\Phi^I$ and their derivatives.

The renormalization of $\Phi^I$ must be done in two steps.
First, we must renormalize the field linearly as
\eqn\elinear{[\Phi^I] = \Phi^I + {\g \over \ep}
\big\lbrace a_1 (\partial_\mu \Phi^J \partial_\mu \Phi^J) \Phi^I +
a_2 \partial^2 \Phi^I \big\rbrace + \dots~,}
where $a_1, a_2$ are finite constants, and we have suppressed the
terms with more than two derivatives.  The above linear
renormalization does not remove all the UV divergences of the
Fourier transforms of the correlation functions.  We must also
subtract the UV
singularities corresponding to two or more fields at the same
point in space.  At one-loop level we
only need counterterms for the product of two $\Phi$'s\foot{At
higher orders of the loop expansion we need counterterms for the
products of more than two fields.}:
\eqna\ePhi
$$
\eqalignno{&\vev{[\Phi^I (p_1) \Phi^J (p_2)]} =
\vev{[\Phi^I] (p_1) [\Phi^J] (p_2)} \cr
&\qquad + \g^2 {1 \over \ep}
\left( b_1 \delta^{IJ} + b_2 \vev{\Phi^I \Phi^J - {\delta^{IJ}\over N}}\right)
(2\pi)^D \delta^{(D)} (p_1 + p_2) &\ePhi a\cr
&\vev{[\Phi^I (p_1) \Phi^J (p_2) \Phi^K (p_3) \Phi^L (p_4)]}^c \cr
& =
\vev{[\Phi^I](p_1) [\Phi^J](p_2) [\Phi^K](p_3) [\Phi^L](p_4)}^c
&\ePhi b\cr
&\qquad + \g^2 {b_2 \over \ep} \left(
\vev{(\Phi^I \Phi^J)(p_1 + p_2) \Phi^K (p_3) \Phi^L (p_4)}^c +
(\hbox{5~other~terms}) \right)~,
\cr}
$$
where the correlation functions are given in the momentum space.
The correlation functions of more number of $\Phi$'s are renormalized
analogously.  The finite constants $b_{1,2}$ are determined to remove
the remaining one-loop divergences.  Note that in the coordinate space
these counterterms would be proportional to delta functions.

We have introduced the seven renormalization constants
$z^{(1,2,3)},~a_{1,2},~b_{1,2}$ to remove the one-loop UV divergences.
(See Appendix A for an explicit expression of the one-loop UV
divergences.)  We can observe that not all the renormalization
constants are independent.  This is a consequence of equations of
motion.  Equations of motion have been discussed extensively in the
literature, and we give a short summary in Appendix B.  For our
purposes we need the following two equations of motion, derived with
only the leading two-derivative term in the lagrangian \enls.

The first equation of motion
\eqn\eEOMab{\eqalign{&\vev{
\left\lbrace (\partial \Phi^I \partial \Phi^I) \Phi^I
+ \partial^2 \Phi^I \right\rbrace (x)
\Phi^{I_1} (x_1) \dots \Phi^{I_n} (x_n)} \cr
&\quad = \sum_{k=1}^n \delta^{(D)} (x-x_k) \vev{
\Phi^{I_1} (x_1) \dots \g \left(- \delta^{II_k} + \Phi^I \Phi^{I_k}\right)
(x_k) \dots \Phi^{I_n} (x_n)}\cr}}
implies that the correlation functions of $\Phi$'s are invariant
under the shift
\eqn\eshiftab{\Delta a_1 = \Delta a_2 = \delta~,\quad
\Delta b_1 = 2(1-1/N) \delta~,\quad \Delta b_2 = - 2 \delta~.}
Similarly, the second equation of motion
\eqn\eEOMza{\eqalign{&- \int d^D x \vev{\left\lbrace
- (\partial \Phi^I \partial \Phi^I)^2 + \partial^2 \Phi^I
\partial^2 \Phi^I \right\rbrace (x) \Phi^{I_1} (x_1) \dots \Phi^{I_n} (x_n)}\cr
&\quad = \sum_{k=1}^n \vev{\Phi^{I_1} (x_1) \dots
\g \left\lbrace (\partial \Phi^I \partial \Phi^I)\Phi^{I_k}
+ \partial^2 \Phi^{I_k} \right\rbrace (x_k) \dots \Phi^{I_n} (x_n)}\cr}}
implies the invariance of the correlation functions under
\eqn\eshiftza{\Delta z^{(1)} = - \Delta z^{(3)} = \Delta a_1 = \Delta a_2~.}
To remove the ambiguities \eshiftab, \eshiftza, it is convenient to
adopt a convention
\eqn\econvention{z^{(3)} = b_2 = 0~.}

Under the above convention \econvention, we can determine the renormalization
constants uniquely as
\eqn\erencons{\eqalign{
&z^{(1)} = {1 \over (4\pi)^2}~\left({1 \over 3} + {N-4\over 2}\right)~,\quad
z^{(2)} =  {1 \over (4\pi)^2}~{2 \over 3}~,\cr
&a_1 = {1 \over (4\pi)^2}~\left({1 \over 2} + {N-4\over 2}\right)~,\quad
a_2 = {1 \over (4\pi)^2}~\left(- {3 \over 2} - {N-4\over 2}\right)~,\cr
&b_1 = 2 a_2~.\cr}}
The two- and four-point functions of $\Phi^i~(i=1,\dots,N-1)$ are now
given by
\eqna\eoneloopNL
$$
\eqalignno{
&{1 \over \g} \vev{[\Phi^i (p_1) \Phi^j (p_2)]} \simeq
\delta^{ij} {1 \over p_1^2} \left( 1 - {2 g_{-4}^{(3)}
\over \g}~ p_1^2 \right) ~,
&\eoneloopNL a\cr
&{1 \over \g^2}
\vev{[\Phi^i (p_1) \Phi^j (p_2) \Phi^k (p_3) \Phi^l (p_4)]}^c\cr
&\quad \simeq
{1 \over \g^2} \vev{[\Phi^i (p_1) \Phi^j (p_2) \Phi^k (p_3)
\Phi^l (p_4)]}_{\hbox{1-loop~MS}}^c \cr
&\qquad + {1 \over p_1^2 p_2^2 p_3^2 p_4^2} \Bigg[
\delta^{ij} \delta^{kl} \Big\lbrace - 8 (g_{-4}^{(1)} + g_{-4}^{(3)})
(p_1 p_2)(p_3 p_4) \cr
&\qquad\qquad - 4 g_{-4}^{(2)}
( (p_1 p_3)(p_2 p_4) + (p_1 p_4)(p_2 p_3) ) &\eoneloopNL b\cr
&\qquad\qquad + 2 g_{-4}^{(3)} (p_1^2 + p_2^2)(p_3^2 + p_4^2) \Big\rbrace
+ \hbox{(t,u-channels)} \Bigg]~,\cr}
$$
where we have suppressed the factor $(2\pi)^4 \delta^{(4)} (p_1 +\dots)$
of momentum conservation.
The first term on the right-hand side of eqn.~\eoneloopNL{b}\ stands for the
correlation function up to one-loop in the minimal subtraction
(MS) scheme in which we take $g_{-4}^{(i)} = 0$.  For a later convenience
we introduce a notation $\Gamma_s^{NL}$ by
\eqn\edefGamma{\eqalign{
&{1 \over \g^2} \vev{[\Phi^i (p_1) \Phi^j (p_2) \Phi^k (p_3)
\Phi^l (p_4)]}_{\hbox{1-loop~MS}}^c \cr
& = {1 \over p_1^2 p_2^2 p_3^2 p_4^2}
\Big[ ~\delta^{ij} \delta^{kl} \left( - \g~s -
\Gamma_s^{NL} (\g; p_1,\dots,p_4) \right)
+ (\hbox{t,u-channels})~\Big]~.\cr}}

We note that the four-point function is free of UV divergences without the
$b_1$ counterterm, but the two-point function would have a UV
divergence proportional to a delta function in space.  The
necessity of the $b_1$ counterterm has been missed in the
previous literature.

Before concluding this section, we note the
inhomogeneous RG equation satisfied by the correlation
functions:
\eqna\ePhiRG
$$
\eqalignno{&\left( \dt + 8 \right)
\vev{[\Phi^I (p_1) \Phi^J (p_2)]} = \g \Big(
\vev{\left\lbrace a_1 (\partial \Phi \partial \Phi)\Phi^I +
a_2 \partial^2 \Phi^I\right\rbrace \Phi^J} +
(I \leftrightarrow J) \Big)\cr
&\qquad\qquad\qquad + b_1 \g^2 \delta^{IJ}
(2 \pi)^4 \delta^{(4)} (p_1+p_2)~,&\ePhiRG a\cr
&\left(\dt + 16\right)
\vev{[\Phi^I (p_1) \Phi^J (p_2) \Phi^K (p_3)
\Phi^L (p_4)]}^c \cr
&= \g \left(
\vev{\left\lbrace a_1 (\partial \Phi \partial \Phi)\Phi^I +
a_2 \partial^2 \Phi^I\right\rbrace
\Phi^J \Phi^K \Phi^L}^c + (\hbox{3~other~terms}) \right)~,\cr
& &\ePhiRG b\cr}
$$
where $a_{1,2},~b_1$ are given by \erencons.

\newsec{The low energy expansion of the linear sigma model}

The O(N) linear sigma model is defined by the lagrangian
\eqn\els{\L = {1 \over 2} Z \partial_\mu \phi^I \partial_\mu \phi^I
+ {Z_\lambda \lambda \over 8} \left(Z \phi^I \phi^I -
{Z_m \over Z_\lambda} v^2 \right)^2~,}
where the elementary fields $\phi^I~(I=1,\dots,N)$ are unconstrained, and
we have written the renormalization constants
$Z, Z_\lambda, Z_m$ explicitly.
A mass squared parameter is defined by
\eqn\emsquared{m^2 \equiv \lambda v^2~.}
It is convenient to adopt the MS scheme to
determine the renormalization constants.  At one-loop we find
\eqn\eZ{Z \simeq 1~,\quad Z_\lambda \simeq 1 +
{\lambda \over (4\pi)^2 \ep}~(N+8)~,\quad
Z_m \simeq 1 + {\lambda \over (4\pi)^2 \ep}~(N+2)~. }
These imply the following one-loop RG equations:
\eqn\eRGlinear{\dt \phi^I = \phi^I~,\quad
\dt \lambda = - {N+8 \over (4\pi)^2}~\lambda^2~,\quad
\dt m^2 = \left( 2 - {N+2 \over (4\pi)^2}~\lambda \right) m^2~.}
For a later convenience, we note that
\eqn\eRGv{\dt {\lambda \over m^2} = \left( -2 - {6 \over (4\pi)^2}~\lambda
\right) {\lambda \over m^2}~.}

We choose the vacuum such that
\eqn\evevL{\vev{\phi^i} = 0\quad(i = 1,\dots,N-1)~,}
corresponding to the convention \evevNL.

Our goal is to obtain the two- and four-point functions of the
elementary fields $\phi^i~(i=1,\dots,N-1)$ up to one-loop.
Especially we wish to obtain their low energy approximations so
that we can compare them with the correlation functions obtained
in the previous section.

At tree-level the four-point vertex (in an obvious notation) is given by
\eqn\etree{\eqalign{- \Gamma_{\hbox{tree}}^{(4)} (p_1,\dots,p_4) &= \lambda
\left[\delta^{ij} \delta^{kl} \left( {m^2 \over s+m^2} -1\right)
+ (\hbox{t,u-channels}) \right]\cr
&\simeq \lambda \left[
\delta^{ij} \delta^{kl} \left( - {s \over m^2} + \left({s \over m^2}\right)^2
\right) + (\hbox{t,u-channels}) \right]~,\cr}}
where $s \equiv (p_1+p_2)^2$, and we have approximated the vertex up
to four derivatives.  The above gives the relation
\eqn\elowest{\g \simeq {\lambda \over m^2} = {1 \over v^2}}
at lowest order in $\lambda$.

The one-loop calculations are straightforward and can be found elsewhere
(for example, \rab\ray).  Up to one-loop the two-point function
of $\phi^i~(i=1,\dots,N-1)$ is given by
\eqn\eonelooptwo{\vev{\phi^i (p_1) \phi^j (p_2)} =
{\delta^{ij} \over p_1^2} \left( 1 - {\lambda \over (4 \pi)^2}
{1 \over 2} + {\lambda \over m^2} {1 \over (4\pi)^2} {1 \over 6}~p_1^2
+ {\rm O} \left( (p_1^2)^2 \right) \right)~,}
where as usual we have suppressed the factor $(2\pi)^4 \delta^{(4)}
(p_1+p_2)$ of momentum conservation.

The one-loop calculation of the four-point function of $\phi^i~(i=1,\dots,
N-1)$ is more elaborate.  But the calculation can be simplified by comparing
the one-loop corrections in the linear sigma model with those
in the non-linear sigma model.  As an example,
let us consider the one-loop contribution to the vertex from
the Feynman diagrams in Fig.~1.  (The solid line corresponds to the
field $\phi^i~(i=1,\dots,N-1)$, and the broken line corresponds
to the field $\sigma \equiv \phi^N - v$.)
\vskip .2in
\centerline{\epsfxsize=.9\hsize \epsfbox{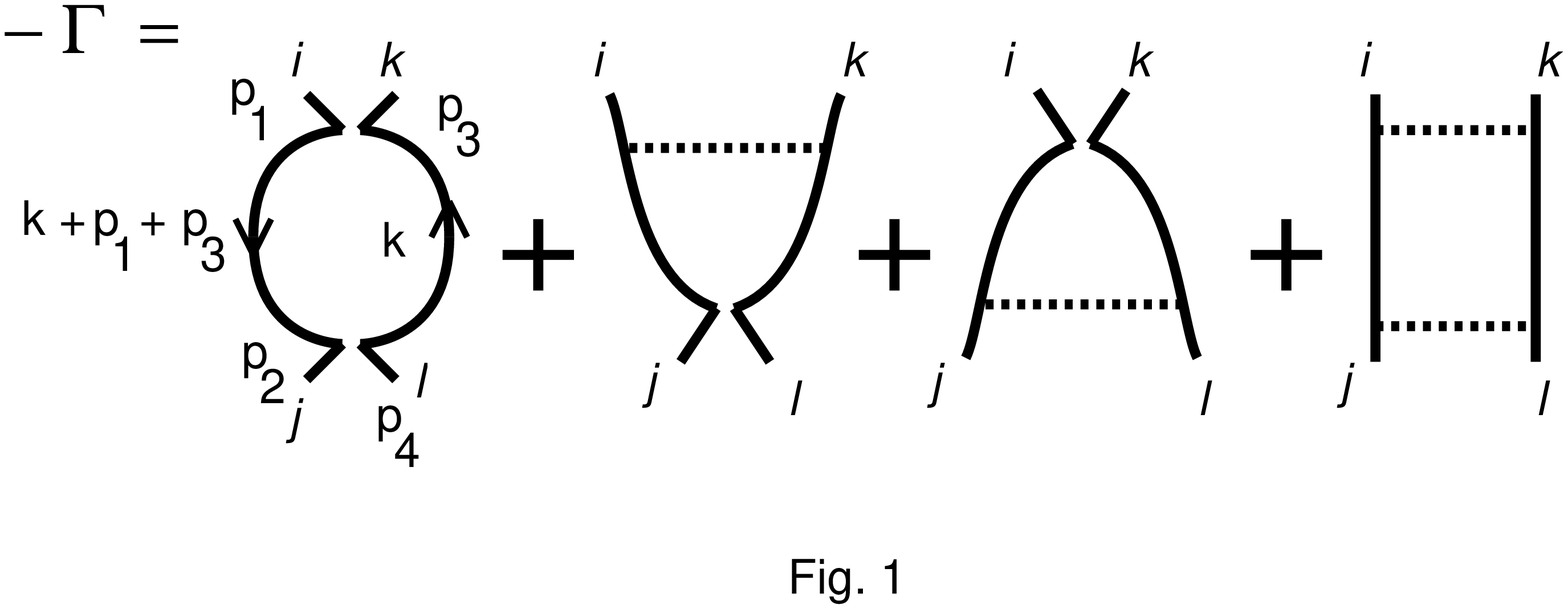}}

\no
We easily find
\eqn\efigone{\eqalign{&- \Gamma = \delta^{ij} \delta^{kl} \lambda^2 \int_k
{1 \over k^2 (k+p_1+p_3)^2} \Bigg[ ~~1
- {m^2 \over m^2 + (k+p_3)^2} \cr
&\qquad - {m^2 \over m^2 +
(k-p_4)^2} + {m^2 \over m^2 + (k+p_3)^2} \cdot
{m^2 \over m^2 + (k-p_4)^2} ~\Bigg]\cr
&\quad= \delta^{ij} \delta^{kl} \left(
G + {\lambda^2 \over m^4}~I \right)~,\cr}}
where
\eqn\eG{G \equiv {\lambda^2 \over m^4} \int_k {(k+p_3)^2 (k-p_4)^2\over
k^2 (k+p_1+p_3)^2}}
corresponds to the contribution of the one-loop graph (Fig.~2) in the low
energy effective theory with $\g = {\lambda \over m^2}$, and
\eqn\eI{I \equiv \int_k
{(k+p_3)^2 (k-p_4)^2 \over k^2 (k+p_1+p_3)^2}
\left[ {m^2 \over m^2 + (k+p_3)^2} {m^2 \over m^2 + (k-p_4)^2}
 - 1 \right]~.}
As far as the contributions up to fourth order in external momenta are
concerned, all the non-local dependence is contained in $G$,
and the integral $I$ is local, i.e., I is a polynomial of
the external momenta.
By expanding the integrand of $I$ in external momenta, we can obtain a low
energy approximation of $I$ without encountering infrared divergences.\foot{
The same technique was used in a proof of the decoupling of massive modes.
See Chapter 8 of Collins's textbook \rcollins.}
\vskip .1in
\centerline{\epsfxsize=.4\hsize \epsfbox{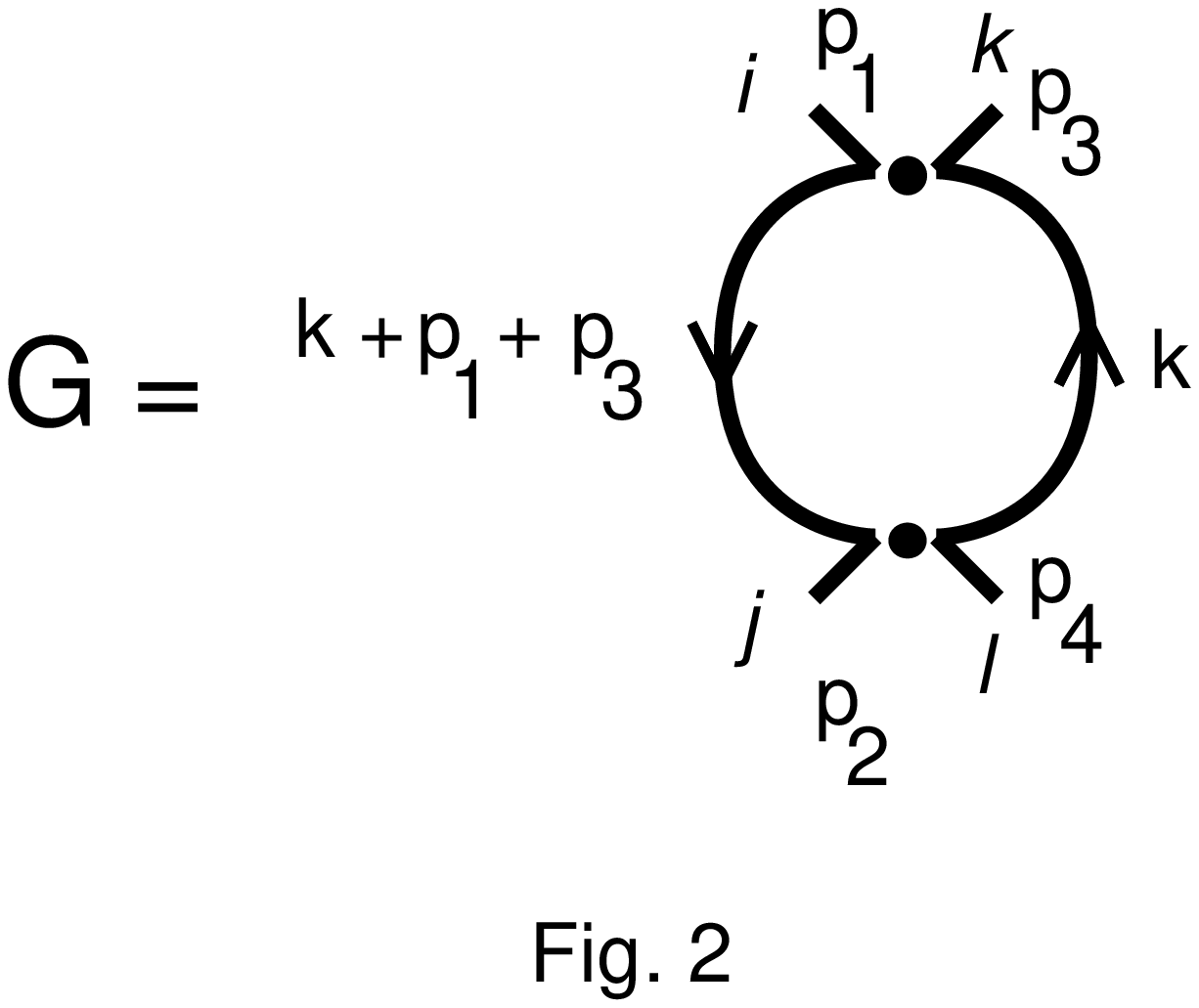}}

The low energy approximations of the other graphs can be evaluated
analogously.  The final result is
\eqnn\eoneloopGamma
$$
\eqalignno{&- \Gamma_{\hbox{1-loop}}^{(4)} (p_1,\dots,p_4)
\simeq \delta^{ij} \delta^{kl} \Bigg[
- \Gamma_s^{NL} \left( {\lambda \over m^2}; p_1,\dots,p_4\right) \cr
&\quad + {\lambda^2 \over m^4} {1 \over (4\pi)^2}
 \bigg\lbrace
- m^2 s \left(3 \ln - {5 \over 2}\right)
+ \left( {4 \over 3}~\ln - {82 \over 9} \right) (p_1 p_2)(p_3 p_4) \cr
&\qquad + \left( {4 \over 3}~\ln - {34 \over 9} \right)
\left( (p_1 p_3)(p_2 p_4) + (p_1 p_4)(p_2 p_3) \right) \cr
&\qquad + \left( \ln - {23 \over 6} \right) s (p_1^2 +\dots+ p_4^2)
&\eoneloopGamma\cr
&\qquad + \left( - {1 \over 2}~\ln + {17 \over 18} \right)
(p_1^2 + p_2^2)(p_3^2 + p_4^2) ~\bigg\rbrace \Bigg] +
(\hbox{t,u-channels}) ~,\cr}
$$
where $\Gamma_s^{NL}$ is defined in eqn.~\edefGamma, and
\eqn\eln{\ln \equiv \ln {m^2 \e^{\gamma} \over 4 \pi}~.}
We have only kept the terms up to fourth order in external momenta.
It is trivial to construct the connected four-point correlation
function $\vev{\phi^i (p_1) \dots \phi^l (p_4)}^c$ from
eqs.~\eonelooptwo, \eoneloopGamma.

\newsec{Off-shell equivalence}

We now wish to construct the correlation functions in the non-linear
sigma model which reproduce the low energy approximations of
the correlation functions in the linear sigma model.

The O(N) symmetry demands that
the elementary field $\phi^I$ of the linear sigma model
corresponds to a linear combination of the fields which
transform as vectors of $O(N)$ in the non-linear sigma model:
\eqn\ephi{\eqalign{\phi_{eff}^I = z(\lambda,m^2) \Big\lbrace
[\Phi^I] +&
\g A_1 (\lambda,m^2) (\partial_\mu \Phi^J \partial_\mu \Phi^J)
\Phi^I\cr
&\quad
+ \g A_2 (\lambda, m^2) \partial^2 \Phi^I + \dots\Big\rbrace~.\cr}}
To assure the off-shell equivalence we also need finite counterterms
which correspond to products of two $\Phi$'s at the same point in space.
Thus, the off-shell equivalence between the two theories means that
we can choose the nine parameters $\g$, $g_{-4}^{(1,2,3)}$,
$z$, $A_{1,2}$, and $B_{1,2}$ such that
\eqna\eequiv
$$
\eqalignno{&\vev{\phi^I (p_1) \phi^J (p_2)} =
\vev{\left[\phi_{eff}^I (p_1) \phi_{eff}^J (p_2)\right]}
+ (2\pi)^4 \delta^{(4)} (p_1+p_2)\cr
&\quad \times \g^2 \left\lbrace \delta^{IJ}  B_1 (\lambda,m^2)
+ B_2 (\lambda, m^2) \vev{\Phi^I \Phi^J - {\delta^{IJ} \over N}}
\right\rbrace  ~, &\eequiv a\cr
& \vev{\phi^I (p_1)\phi^J (p_2) \phi^K (p_3) \phi^L (p_4)}^c =
\vev{\left[\phi_{eff}^I (p_1) \phi_{eff}^J (p_2)
\phi_{eff}^K (p_3) \phi_{eff}^L (p_4)\right]}^c\cr
&\quad + \g^2 B_2 (\lambda, m^2) \Big\lbrace
\vev{(\Phi^I \Phi^J)(p_1+p_2) \Phi^K (p_3) \Phi^L (p_4)}^c
+ (\hbox{5~other~terms}) \Big\rbrace~.\cr
& &\eequiv b\cr}
$$
As in sect.~2, the equations of motion \eEOMab, \eEOMza\ imply
that only seven of the nine parameters are independent, and
following \econvention\ we adopt the convention
\eqn\econventionII{g_{-4}^{(3)} = B_2 = 0~.}

{}From the results of the previous two sections, we obtain
the following results to order $\lambda^2$:
\eqna\ematching
$$
\eqalignno{
\g &\simeq {\lambda \over m^2}
\left( 1 + {\lambda \over (4 \pi)^2}
\left( 3 \ln - {7 \over 2} \right) \right) &\ematching a\cr
g_{-4}^{(1)} &\simeq - {1 \over 2} {\lambda \over m^4}
+ {\lambda^2 \over m^4} {1 \over (4 \pi)^2}
\left( - {1 \over 6} \ln + {41 \over 36} \right)&\ematching b\cr
g_{-4}^{(2)} &\simeq {\lambda^2 \over m^4} {1 \over (4\pi)^2} \left(
- {1 \over 3} \ln + {17 \over 18} \right)&\ematching c\cr
\g^2 A_1 &\simeq - {1 \over 2} {\lambda \over m^4} +
{\lambda^2 \over m^4} {1 \over (4 \pi)^2}
\left( - {1 \over 4} \ln + {17 \over 36} \right)&\ematching d\cr
\g^2 A_2 &\simeq {1 \over 2} {\lambda \over m^4} + {\lambda^2 \over
m^4} {1 \over (4 \pi)^2} \left(
{3 \over 4} \ln - {127 \over 36} \right)&\ematching e\cr
\g^2 B_1 &\simeq {\lambda \over m^4} + {\lambda^2 \over m^4}
{1 \over (4 \pi)^2} \left( {3 \over 2} \ln - {62 \over 9} \right)~,
&\ematching f\cr}
$$
where $\ln$ is defined by eqn.~\eln.  The normalization constant $z$
is given, to order $\lambda$, by
\eqn\ez{z \simeq 1 - {\lambda \over (4\pi)^2} {1 \over 4}~.}

Let us note the consistency between the dependence of $\ln m^2$ in the
above results and the one-loop RG.  (The one-loop RG equations for the
non-linear sigma model are given by eqs.~\egfourRG\ and \ePhiRG{}.
Those for the linear sigma model are given by eqs.~\eRGlinear, \eRGv.)
Since the anomalous dimension of the field $\phi^I$ vanishes to order
$\lambda$, the normalization constant $z$ should be independent of the
$\ln m^2$, which is consistent with eqn.~\ez.  Because of eqn.~\eRGv,
eqn.~\ematching{a}\ is consistent with
\eqn\eRGg{\dt \g = -2 \g~.}
Eqs.~\ematching{b,c}\ give
\eqn\eRGgfour{\eqalign{ \left( \dt + 4
\right) g_{-4}^{(1)} &\simeq - {\g^2 \over (4\pi)^2} \left( {1 \over
3} + {N-4 \over 2} \right) = - z^{(1)} \g^2 \cr \left(\dt + 4 \right)
g_{-4}^{(2)} &\simeq - {\g^2 \over (4\pi)^2} {2 \over 3} = - z^{(2)}
\g^2~,\cr}}
which agree with eqn.~\egfourRG\ due to eqs.~\erencons.
The constants $A_{1,2},~B_1$ satisfy
\eqn\eRGAB{\eqalign{\left( \dt
+ 4 \right) \g^2 A_1 &\simeq - {\g^2 \over (4\pi)^2} \left( {1 \over
2} + {N-4\over 2} \right) = - a_1 \g^2\cr \left( \dt + 4 \right) \g^2
A_2 &\simeq {\g^2 \over (4\pi)^2} \left({3 \over 2} + {N-4 \over 2}
\right) = - a_2 \g^2\cr \left( \dt + 4 \right) \g^2 B_1 &\simeq {\g^2
\over (4\pi)^2} \big( 3 + (N-4) \big) = - b_1 \g^2~.\cr}}
Hence, the inhomogeneous counterterms given by $A_{1,2},~B_1$ cancel
the inhomogeneous terms in the RG eqs.~\ePhiRG{}.

\newsec{Concluding remarks}

In this note we have shown how to achieve the off-shell equivalence
between the O(N) linear and non-linear sigma models by choosing the
parameters of the non-linear sigma model as appropriate functions of
$\lambda$ and $m^2$ of the linear sigma model.  At one-loop level we
have determined the seven parameters $\g$, $g_{-4}^{(1,2)}$, $z$,
$A_{1,2}$, and $B_1$ in terms of $\lambda,~m^2$ as in
eqs.~\ematching{}, \ez.

The emphasis of this note is that the off-shell equivalence can be
achieved by introducing all possible counterterms allowed by the O(N)
invariance.  We especially emphasize the importance of the inhomogeneous
counterterm proportional to $B_1$ in eqn.~\eequiv{a}.
At higher loop orders we need to
introduce more inhomogeneous counterterms corresponding to
products of more than two fields at the same point in space.
Their role will increase at higher orders.

Another point of emphasis is the irrelevance of any particular way of
parametrizing the fields $\Phi^I$ which satisfy the non-linear
constraint \enonlinear.  We only need to be concerned with the linear
representations of O(N), but not with their non-linear realizations.

Out of the seven parameters of the non-linear sigma model, only three,
i.e, $\g$ and $g_{-4}^{(1,2)}$, affect the S-matrix elements.  (See
ref.~\rarzt\ for a recent discussion.)  As is well known, the S-matrix
elements are independent of the choice of a particular interpolating
field.  Hence, the parameters $z$ and $A_{1,2}$ are irrelevant.  The
counterterm proportional to $B_1$ vanishes when the external legs are
amputated, and the momenta are put on the mass shell.

\vskip .1in
\no
{\bf Acknowledgment}: The present work was done while the
author was visiting the Theory Group of KEK in Japan.  He would
like to thank the group for a stimulating atmosphere.

\vfill\eject

\appendix{A}{One-loop calculations in the O(N) non-linear sigma model}

For completeness we give the results of the one-loop
calculations in the O(N) non-linear sigma model \rab\ray.  The
two- and four-point functions of $\Phi^i~(i=1,\dots,N-1)$ up to
one-loop level are given as follows:
\eqna\eoneloop
$$
\eqalignno{
&{1 \over \g} \vev{[\Phi^i (p_1) \Phi^j (p_2)]} \cr
&\quad =
{\delta^{ij} \over p_1^2} \left[ 1 +
(p_1^2)^2 \left\lbrace
- 2 G_{-4}^{(3)} +
{\g^2 \over \ep} \left( - 2 a_2 + b_1 - {b_2 \over N} \right)
\right\rbrace\right]&\eoneloop a\cr
&{1 \over \g^2} \vev{[\Phi^i (p_1) \Phi^j (p_2) \Phi^k (p_3)
\Phi^l (p_4)]}^c \cr
&\quad = {1 \over \g^2} \vev{\Phi^i (p_1) \Phi^j (p_2) \Phi^k
(p_3)
\Phi^l (p_4)}_{\hbox{1-loop~MS}}^c + {1 \over p_1^2 p_2^2 p_3^2
p_4^2} \Bigg\lbrace\cr
&\qquad \delta^{ij} \delta^{kl} \Bigg[\,\,
\left\lbrace {\g^2 \over (4\pi)^2 \ep} \left({8 \over 3} + 4(N-4) \right)
- 8 \left(G_{-4}^{(1)} + G_{-4}^{(3)}\right)
\right\rbrace (p_1 p_2)(p_3 p_4)\cr
&\qquad\quad
+ \left( {\g^2 \over (4\pi)^2 \ep} {8 \over 3} - 4 G_{-4}^{(2)} \right)
\big( (p_1 p_3)(p_2 p_4) + (p_1 p_4)(p_2 p_3) \big) &\eoneloop b\cr
&\quad
 + \left\{ {\g^2 \over \ep} \left( - {1 + (N-4) \over (4\pi)^2}
+ 2 a_1 + b_2 \right) + 2 G_{-4}^{(3)} \right\}
(p_1^2 + p_2^2)(p_3^2 + p_4^2)\cr
&\quad
+ {\g^2 \over \ep} \left( {2 + (N-4) \over (4\pi)^2} - a_1 + a_2
\right) s (p_1^2 +\dots+ p_4^2) \,\, \Bigg] + (\hbox{t,u-channels})
\Bigg\rbrace~,\cr}
$$
where we have suppressed the factor $(2\pi)^D \delta^{(D)} (p_1 +\dots)$
corresponding to the overall momentum conservation, and
$s \equiv (p_1+p_2)^2$.  The first term on the right-hand side
of eqn.~\eoneloop{b}\ denotes the one-loop correlation function
in the MS scheme.  The above results
are invariant under the shifts \eshiftab, \eshiftza\ as should be.

\appendix{B}{Equations of motion}

Equations of motion in the context of low energy effective
theories has been discussed recently in ref.~\rarzt.
We will sketch a derivation of the equations of motion which
are necessary in the main text.

Let $T^I (\Phi(x))$, a local function of the field $\Phi^I$ and
its derivatives, be a vector of O(N).  For example, we can take
\eqn\evector{\eqalign{T^I (\Phi) &= a_0 \Phi^I
+ a_{1,1} (\partial_\mu \Phi^J \partial_\mu \Phi^J) \Phi^I
+ a_{1,2} \partial^2 \Phi^I \cr
&\quad + {\rm terms~with~4~or~more~derivatives}~.\cr}}
We consider an infinitesimal change of variable
\eqn\echange{\Phi^I \to {\Phi'}^I = \Phi^I + \ep ~T^I (\Phi)~.}
Since ${\Phi'}^I$ must be a unit vector, we must find
\eqn\ePhiT{\Phi^I T^I = 0~.}
The most general solution to this constraint is given by
\eqn\eTsol{T^I (\Phi) = (\delta^{IJ} - \Phi^I \Phi^J) \tilde{T}^J (\Phi)~,}
where $\tilde{T}^I$ is an arbitrary vector.  For example, we have
\eqn\eTexample{T^I (\Phi) = a \left(\partial^2 \Phi^I +
(\partial_\mu \Phi^J \partial_\mu \Phi^J) \Phi^I \right) +
{\rm 4~or~more~derivatives}~.}

Let $S [\Phi]$ be an O(N) invariant action.  For a vector $T^I (\Phi)$
which satisfies the condition \ePhiT, we find
\eqn\edS{S[\Phi + \ep ~T] - S[\Phi] = \int d^D x~\ep (x)
\, T^I (\Phi(x))\, {\delta S \over
\delta \Phi^I (x)} \Big\vert_{n.c.}~,}
where we take $\ep (x)$ to be an arbitrary infinitesimal
function, and $n.c.$ stands for no constraint, i.e., we do not
take into account that $\Phi^I$ is a unit vector when we take
the derivative.  For example, for
\eqn\eSexample{S = \int d^D x~{1 \over 2 \g} \partial_\mu \Phi^I
\partial_\mu \Phi^I~,}
we obtain
\eqn\encexample{{\delta S \over \delta \Phi^I (x)}\Big\vert_{n.c}
= - {1 \over \g} \partial^2 \Phi^I~.}

Now we consider the correlation function
\eqn\ecorrelation{\vev{T_1^{I_1} (\Phi(x_1)) \dots T_n^{I_n} (\Phi(x_n))}
\equiv \int [d\Phi] ~T_1^{I_1} (\Phi(x_1)) \dots T_n^{I_n} (\Phi(x_n))
\exp [ - S [\Phi]]~,}
where $T_i^I$ are arbitrary O(N) vectors, not necessarily
satisfying the constraint \ePhiT.  If we use the dimension
regularization, the jacobian of the infinitesimal field
transformation \echange\ is unity.  Hence, we obtain the
following equation of motion:
\eqn\eEOM{\eqalign{&\int d^D x ~\ep (x) \vev{T^I (\Phi(x))
{\delta S \over \delta \Phi^I (x)} \Big\vert_{n.c.}
T_1^{I_1} (\Phi (x_1)) \dots T_n^{I_n} (\Phi (x_n))}\cr
&= \sum_{k=1}^n \vev{T_1^{I_1} (\Phi (x_1)) \dots
\left\lbrace T_k^{I_k} (\Phi + \ep T(\Phi(x_k))) - T_k^{I_k} (\Phi(x_k))
\right\rbrace \dots T_n^{I_n} (\Phi (x_n))}.\cr}}

For our purposes, it suffices to take the simple
action \eSexample\ and consider two special cases:

\no
(i) $T_i^I (\Phi) = \Phi^I$ and $T^I (\Phi) = \partial^2 \Phi^I
+ (\partial \Phi^J \partial \Phi^J) \Phi^I$:
\eqn\eEOMI{\eqalign{&
\int d^D x~\vev{\big\lbrace - \partial^2 \Phi^I \partial^2 \Phi^I
+ (\partial \Phi^J \partial \Phi^J)^2 \big\rbrace (x)
\Phi^{I_1} (x_1) \dots \Phi^{I_n} (x_n)} \cr
&\quad = \sum_{k=1}^n \vev{\Phi^{I_1} (x_0) \dots \g
\big\lbrace \partial^2 \Phi^{I_k} +
(\partial \Phi^J \partial \Phi^J) \Phi^{I_k}\big\rbrace (x_k) \dots
\Phi^{I_n} (x_n)}\cr}}
This gives eqn.~\eEOMza.

\no
(ii) $T_i^I (\Phi) = \Phi^I$ and $T^I (\Phi) = \delta^{KI} - \Phi^K \Phi^I$
(this is a vector for a fixed $K$):
\eqn\eEOMII{\eqalign{&
\vev{\big\lbrace \partial^2 \Phi^K + (\partial \Phi^J \partial \Phi^J)
\Phi^K \big\rbrace (x) \Phi^{I_1} (x_1) \dots \Phi^{I_n} (x_n)}\cr
&\quad = \sum_{k=1}^n \delta^{(D)} (x-x_k) \vev{\Phi^{I_1} (x_1) \dots
\g \left( - \delta^{KI_k} + \Phi^K \Phi^{I_k} (x_k)\right) \dots
\Phi^{I_n} (x_n)}\cr}}
This gives eqn.~\eEOMab.

\listrefs

\bye